\documentclass[twocolumn,showpacs,preprintnumbers,amsmath,amssymb]{revtex4}
\usepackage{graphicx}
\usepackage{bm}
\begin{document}
\title{ \large SIMULACI\'ON POR AUT\'OMATAS CELULARES DE LAS CONFORMACIONES ESPACIALES
DE POLIELECTROLITOS.}
\author{{\bf Carlos Echeverria$^{1,2}${\footnote{Corresponding author: e-mail: cecheve@ula.ve}}, Wilmer Olivares-Rivas$^{2}$, and  Kay Tucci$^{3}$}}.\\
\affiliation{ $^{1}$Lab. de F\'isica Aplicada y Computacional, Depto. de Matem\'atica y F\'isica, Universidad Nacional Experimental del T\'achira, T\'achira, Venezuela.\\
$^{2}$Grupo de Quimicof\'isica de Fluidos y Fen\'omenos Interfaciales, Depto. de Q\'imica, Fac. de Ciencias, Universidad de Los Andes, M\'erida, Venezuela.\\
$^{3}$SUMA-CeSiMo, Fac. de Ciencias, Universidad de Los Andes  M\'erida, Venezuela.
}
\begin{abstract}
%\vspace{1cm}
\begin{center}RESUMEN\\
\end{center}
{\small {\bf
Hemos llevado a cabo simulaciones del tipo  Aut\'omata Celular para un modelo de un polielectrolito en diluci\'on infinita, para reproducir de manera cualitativa sus propiedades conformacionales. Nuestros resultados predicen las llamadas estructuras de \emph{collar de perlas}, las cuales se comparan bien con simulaciones de Din\'amica Molecular m\'as elaboradas y costosas.}}\\

\vspace{1.5 cm}
\begin{center}
{\large {\bf {CELLULAR AUTOMATA SIMULATION OF THE SPATIAL CONFORMATIONS OF\ POLYELECTROLYTES.}}}
\end{center}
%\vspace{1.0 cm}
 
\begin{center}ABSTRACT\\
\end{center}
{\small {\bf
We carried out a Cellular Automata simulation of a model polyelectrolyte solution at infinite dilution, in order to reproduce qualitatively its conformational properties. Our results predict the so called \emph{pearl necklace} structures, which compare favorably with the more elaborated and costly Molecular Dynamics simulations.}}
\end{abstract}
\vspace{2.0 cm}
\maketitle

\section{Introduction}
%////////////

Polyelectrolyte (PE) solutions are systems widely studied since they show properties that are of fundamental interest for applications in health science, food industry, water treatment, surface coatings, oil industry, among other fields. In fact, one of the problems found in genetic engineering in the appearance of conformational changes of the ADN molecule, which is a charged polyelectrolyte.\cite{YTY1}.\\

    Here we study an infinite dilution polyelectrolyte solution, so that,
the interaction among polyelectrolyte macromolecules are negligible. We model the polyelectrolyte as having dissociable functional groups that give rise to charged sites and counter-ions in aqueous solution. The long range interactions arising from these multiple charges are responsible for their macroscopic complex properties, which can not be explained by regular polymer theories. The spatial structures of these materials in solution have been studied extensively, particularly with a scaling theory\cite{DCR95,DRO1} that are not appropriate for highly charged PE. The first simulations carried out for a single chain predicted the formation of groups of monomers, as the fraction of charged monomers increased. Such structures are known as pearl necklaces. The size of such pearls and the distance between then is determined by the balance between the electrostatic repulsion and
steric effects.\\

These pearl necklace structures have also been found  in Molecular Dynamics (MD) simulations\cite{LH1,MHK1,MK1,LH2,LH3,LHK1,LHK2}. In this paper we are interested in the application of the much simpler Cellular Automata simulation to characterize the main features of a polyelectrolyte that could be responsible for such conformations.
The complete simulation of this complex system requires the description of a model in terms of potential or forces.  In the MD simulations of Limbach and  Holm \cite{LH1}, the monomers are connected along the chain by the finite extendible nonlinear elastic (FENE) bond represented by the potential energy.

\begin{equation}
U_{E}(r) = -\frac{1}{2}K_{ E}R_0^2\ln \left(1-\frac{r^2}{R_0^2} \right)
\end{equation}
where $r$ is the distance between two bonded monomers,
$K_{E}$, is the elastic bond constant, $\sigma$ is the monomer diameter, $k_B$ is  Boltzmann's constant, $T$ is the absolute temperature and the parameter $R_0$ represents the maximum extension of the bond between two neighbor monomers. Two charged sites i and j, with charges $eq_{i}$ and $eq_{j}$, a distance $r_{ij}$ apart, interact with the electrostatic Coulomb potential
\begin{equation}
U_{C}(r_{ij}) = k_B T \frac{\lambda_Bq_iq_j}{r_{ij}}.
\end{equation}
This potential is weighed by the Bjerrum length
$\lambda_B=e^2/[4\pi \epsilon_o\epsilon_sk_BT]$, where
$\epsilon_s$ and $\epsilon_o$ are the solvent permitivity and the vacuum permitivity  respectively, $e$ is the electric charge unit. The parameter $\lambda_B$ is a measure of the strength of the electrostatic force as compared to the kinetic energy. The length ratio $\sigma/\lambda_B$ is a measure of the reduced temperature $ T/[e^2/(4\pi \epsilon_o\epsilon_s k_B\sigma )]$.

The short range and van der Waals  interaction between any two particles or monomers is represented in the MD simulation by a typical truncated Lennard-Jones potential
\begin{equation}
U_{LJ} (r_{ij}) = \left\lbrace
\begin{array}{lll}
4\epsilon \left[ \left( \frac{\sigma}{r_{ij}} \right)^{12} - \left(
\frac{\sigma}{r_{ij}} \right)^{6}
\right] + \epsilon_{cut} & r_{ij} < R_{c},\\
0 &  r_{ij} > R_{c}
\end{array}\right.
\end{equation}
where $\epsilon$ is the potential energy well depth and  $\epsilon_{cut}$ is the cut off energy. This potential prevents the superposition of the bonding monomers. Counter-ions interact via a purely repulsive LJ interaction with
$R_c= 2^{1/6}\sigma$.

\section{Cellular Automata Model}
Even though in the Cellular Automata simulation we do not use any form of potential energies or forces in an explicit manner, the {\it rules} for the movement of the different particles must be inspired on a model defined in terms of such potentials.
We therefore establish our rules based on the essence of the previous three potentials.\\

%FIGURE 1
\begin{figure}
\hbox{\centerline{\includegraphics[scale=0.5]{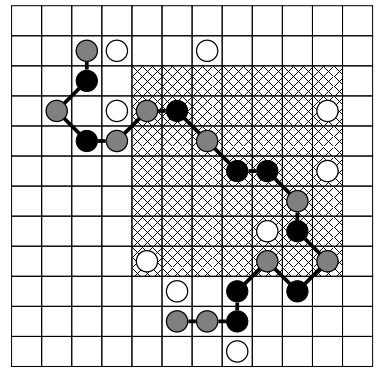}}}
\caption{\em Celular Automata Model. The dark bonded circles represent the charged monomers, the gray circles are the neutral monomers and the open circles are the counter-ions. The shaded area corresponds to the range volume $\lambda_B= 3 \sigma$   }
\end{figure}
\vspace{1cm}
\subsection{Polyelectrolyte Chain Construction Rules}
The polymer is constructed by placing the monomers in a three dimensional cubic network of side $L$ and volume $L^{3}$. Each cell then has $26$ neighbors and represents a monomer with a monovalent charge $+1$, $-1$, or, for a neutral monomer, $0$, as depicted in Fig.(1) in two dimensions. Out of a total of $N$ monomers in the chain, it is assumed that a given fraction $f$ is charged. The polymer is then constructed by randomly binding consecutive sites in the network. Each monomer could be charged or uncharged, with a distribution chosen randomly. A key step on the construction of the polyion is the spatial location of the dissociated counter-ions. We place the counter-ions also randomly in free cells in the volume around the charged monomer within a distance $\lambda_B$, that is, in a volume $(2\lambda_B)^3$ centered on the charged site. The use of the Bjerrum parameter, which is related to the quality of the solvent, ensures the conservation of the total electroneutrality but gives a spatial distribution of counter-ions around the charged sites.\\

So, each monomer $i$ of the system is represented in a $Nx4$ matrix  where each element $M(m_x,m_y,m_z,q_i)$ indicates the $i^{th}$ polymer with charge $q_i$, that could be $-1, +1$ or $0$, at the positions $x,y,z$ given by the cell label $[m_x,m_y,m_z]$. The counter-ions with opposite charge are represented by a similar $fNx4$ matrix $M_C$. \\

For simplicity we chose monomers with dissociable groups that give a site with a positive charge. We then set the following displacement rules for the different particles:\\

\subsection{Displacement Rules}

{\bf \emph{ Neutral Monomer Particle}}
\begin{enumerate}
\item Locate the unoccupied nearest neighbor sites. The new position where a move could be acceptable are those where it does not superimposed with any other particle in the system and where no bond is broken.
\item Count the amount of monomer particles around the current position and around every unoccupied neighbor, within a cube of volume $(2L)^3$ centered on it.
\item Move the test neutral monomer to  the position that has the higher amount of monomers around it, including the current one if it were the case.
\end{enumerate}

{\bf \emph{Positively Charged Monomer Particle }}
\begin{enumerate}
\item As before, locate the unoccupied acceptable nearest neighbor sites.
\item Count the amount of charge around the current position and around every unoccupied neighbor, within a cube of volume $(2\lambda_B)^3$ centered on it.
\item Move the test positive charged monomer to  the position that has the lowest positive charge around it, including the current one if it were the case.
\end{enumerate}

{\bf \emph{Negatively charged counter-ion }}
\begin{enumerate}
\item Move randomly to an unoccupied site within a cube of volume $(2\lambda_B)^3$ centered on the accepted new position of the corresponding positive monomer in the polymeric chain.
\end{enumerate}

%FIGURE 2
\begin{widetext}
\begin{center}
\begin{figure}[ht]
\hbox{\centerline{\includegraphics[scale=1.3]{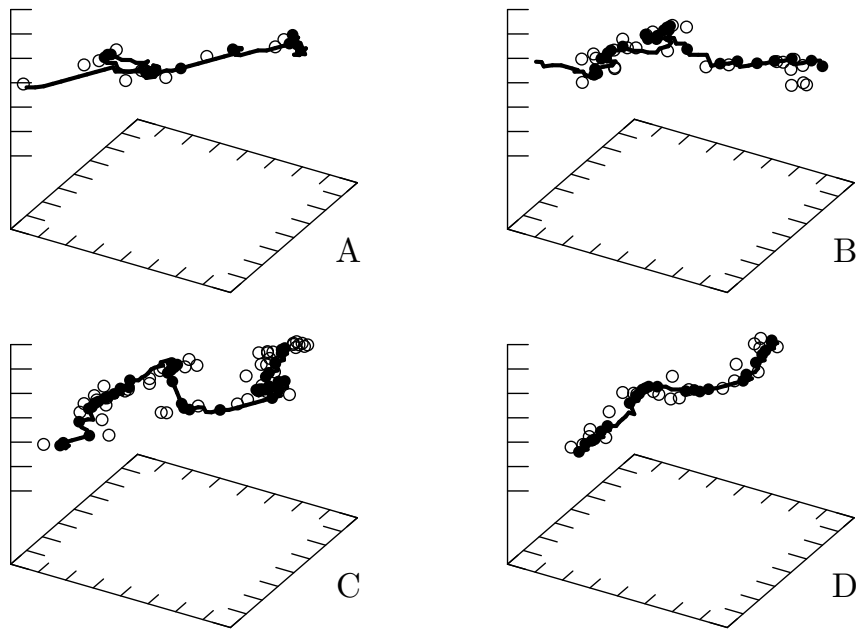}}}
\caption{\em Final polyelectrolyte configuration for different charge fractions
$f$: A) 0.1 B) 0.3 C) 0.5 D) 0.7. The pearl necklace conformations appear as the charge fraction increases.}\vspace{1.0 cm}
\end{figure}
\end{center}
\end{widetext}

\section{RESULTS AND DISCUSSION }
\begin{center}
\begin{figure*}
\hbox{\centerline{\includegraphics[scale=0.45,angle=90]{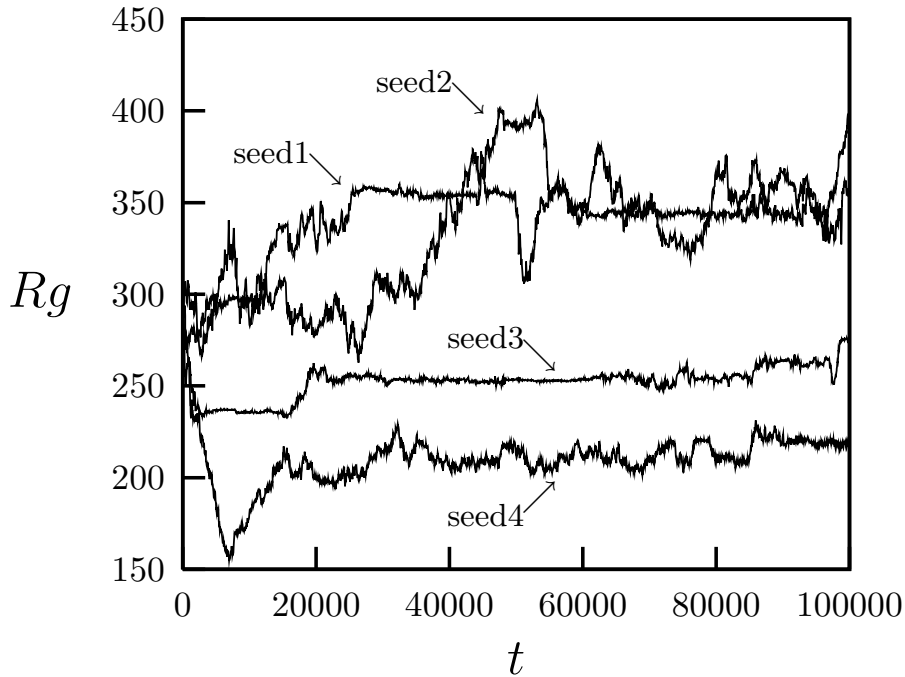}}}
\caption{\em Evolution of the Radius of Gyration $R_g$ as a function of time steps $t$, for  $N=100$, $\lambda_B/\sigma = 3$ and $f=0.5$. Each curve corresponds to different random charged sites distributions, obtained with different seeds.}
\vspace{1.0 cm}
\end{figure*}
\end{center}
We denote the position of monomer i with $r_i$ and the distance
between two particles i and j with $r_ij$. The center of mass for
the chain is then
$R_s =
\frac{1}{N}\sum_{i=1}^N r_i$.
and the center of mass
coordinates are $x_i = r_i - R_s$. A parameter that is useful in the study of the spatial conformations of a polymer is the  radius of gyration $R_G$, defined as
\begin{equation}
R_g^2 = \frac{1}{N}\sum_{i=1}^N x_i^2
\end{equation}

According with our construction and movement rules, we can vary the length of the chain $L$, or the number of monomers $N$, satisfying $L=N\sigma$ and the number of charged monomers, $N_c=fN$. We also take as an independent variable the parameter $\lambda_B$, which determines the number of cells, of size $\sigma$, where the range of the electrostatic attraction between a charged monomer and contra-ion extends. The charge distribution of the sequence of charged and uncharged monomers is determined randomly depending on the initial random seed.\\

%FIGURE 3 Rg vs t vs seed \newpage
%\begin{widetext}
%\begin{center}
%\begin{figure*}
%\hbox{\centerline{\includegraphics[scale=0.4,angle=90]{nadRg.ps}}}
%\caption{\em Evolution of the Radius of Gyration $R_g$ as a function of time steps $t$, for  $N=100$, $\lambda_B/\sigma = 3$ and $f=0.5$. Each curve corresponds to different random charged sites distributions, obtained with different seeds.}
%\vspace{1.0 cm}
%\end{figure*}
%\end{center}
%\end{widetext}
In Fig.(2) we show some of the equilibrium conformations obtained for a polyelectrolyte with $N=100$ monomers, with $\lambda_B=3\sigma$, for several charge fractions. The line represents the polyelectrolyte, the filled black bonded circles  represent the charged monomers. The counter-ions in solution are represented by open non-bonded circles. For clarity, the neutral monomers are not shown. For a low charge fraction of $f=0.1$, the polyion presents an elongated string appearance. As the charge fraction is increased the polyion contracts and some groups  of monomers tend to form  clusters, so that, already for $f=0.5$, it  shows the locally collapsed structures known as pearl necklace. These results are very similar to those obtained by the Molecular Dynamics simulations of Limbach and Holm \cite{DRO1,LH1}\\

It is important to notice that for a given total charge, determined by the fraction $f$, different distributions of the charged monomers give different conformations. To study this behavior, we have carried out simulations with several initial seeds for $N=100$ monomers, with a fixed fraction $f=0.5$ and a range parameter of  $\lambda_B/\sigma = 3$. In Fig.(3), we show the temporal evolution of the radius of gyration $R_g$. In all cases the conformations change from the initial given $R_g$ value to a plateau value that correspond to the equilibrium structures. Fig.(3) clearly show that the plateau $R_g$ values, and thence, the final conformations depend on the charge distribution. In Fig.(4) we show snapshots of final structures corresponding with the different seeds of Fig.(3).\\

%FIGURE 4 Conformations vs seeds (from Fig3)
\begin{widetext}
\begin{center}
\begin{figure}[t]
\hbox{\centerline{\includegraphics[scale=1.4]{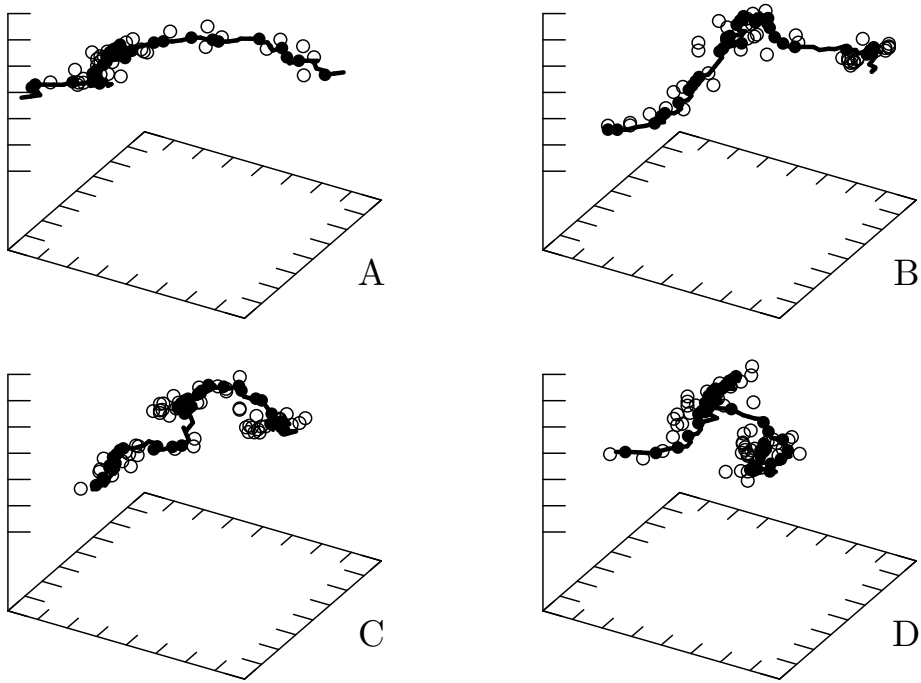}}}
\caption{\em Equilibrium Polyelectrolyte structures in the plateau $R_g$ region for the charge of distributions of Fig.(3). A) seed1, B) seed2, C) seed3 and D) seed4.}
\vspace{1.6 cm}
\end{figure}
\end{center}
\end{widetext}

As we can see from Fig.(4), the formation of the pearl necklace structures is independent of the charge sites distribution, for a given $f$ and $\lambda_B$. These clusters seem to be stabilized by the counter-ions as a consequence of the electroneutrality condition that we force to be satisfied. This is so because in our model the degree of freedom of the mobile counter-ions is much higher that that of the monomers tied to the chain. The strong repulsions that originate by the formation of clusters of neutral and positively charged monomers is compensated by the counter-ion cloud that forms around it.

%FIGURE 5
\begin{widetext}
\begin{center}
\begin{figure}[b]
\hbox{\centerline{\includegraphics[scale=1.0]{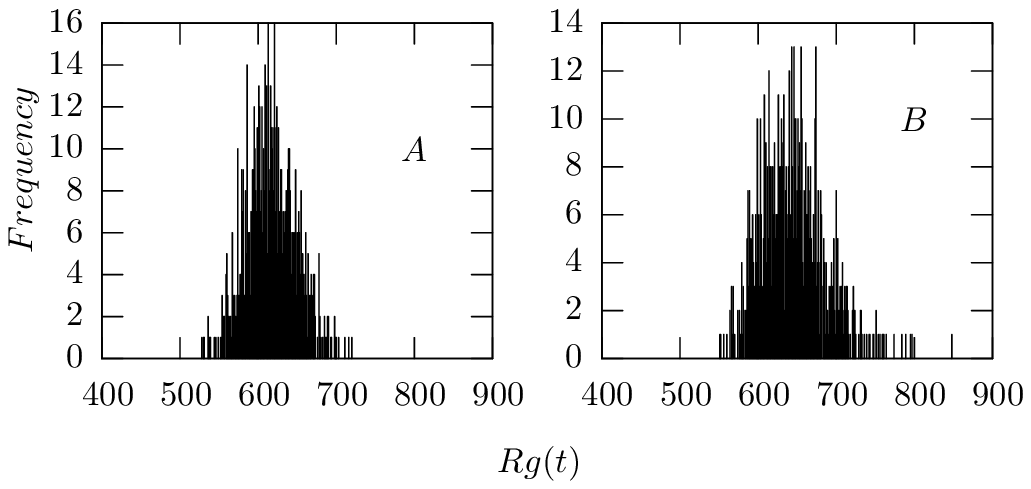}}}
\caption{\em Histograms for $R_g$ for $N=150$, $f=0.5$ and two values of  $\lambda_B/\sigma$,
A) 3 and B)9 }
\vspace{0.0 cm}
\end{figure}
\end{center}

\end{widetext}

In order to study the reproducibility of the configurations found, we carried out a large number of simulation runs, for fixed values of $N=150$, $f=0.5$, $\lambda_B/\sigma=3$ and $9$ and for the same initial charge distribution. In Fig.(5) we show the histograms for the frequency with which a given value of $R_g$ appears. We can see that the distribution of the $R_g$ is very closely a gaussian with a reasonably low dispersion of less than a $10\%$ about the mean value.\\
%\newpage
%FIGURE 6 Structures for lambda
\begin{widetext}
\begin{center}
\begin{figure}[h]
\hbox{\centerline{\includegraphics[scale=1.3]{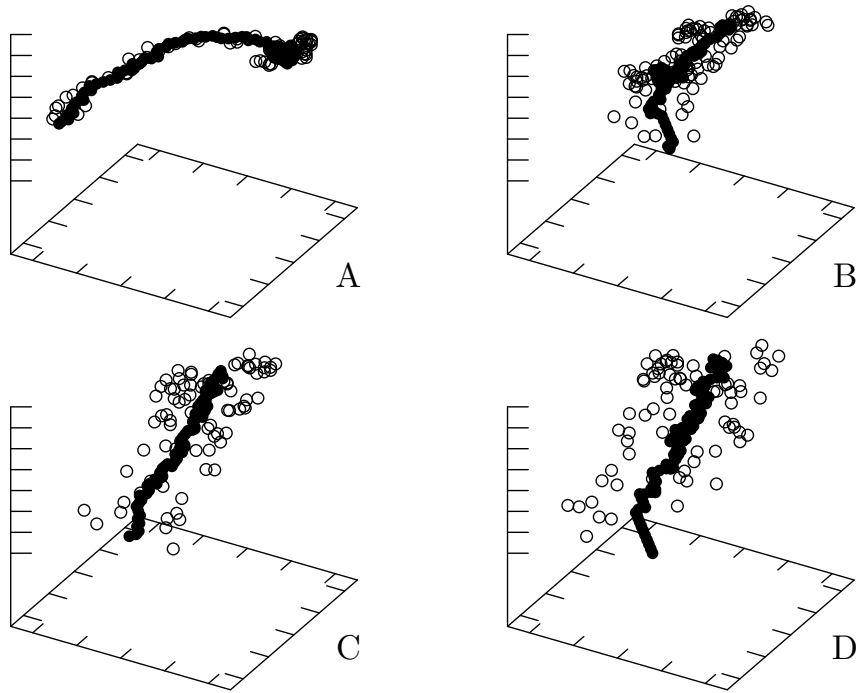}}}
\caption{\em Polyelectrolyte conformations for several values of the reduced Bjerrum parameter $\lambda_B/\sigma$ A) 3,B) 6,C) 9 and D) 12. Here $f=0.9$ and $N=100$ and we used the same initial charge distributions}
\end{figure}
\end{center}
\end{widetext}

In Fig.(5B) we can see that similar distribution is obtained when the range parameter $\lambda_B/\sigma$ increases to a value of $9$. We have further tested the effect of the parameter $\lambda_B$ by generating structures for different values of it. In Fig.(6) we show some equilibrium  conformations for $\lambda_B/\sigma$ equal to  A) 3,B) 6,C) 9 and D) 12. Here we use a large charge fraction $f=0.5$ and $N=100$ and we used the same initial charge distributions in all cases. As we can observe as Bjerrum length increases the final polyelectrolyte structures become more compact. The number or pearls or conglomerates is higher for the lower values of $\lambda_B$, a result similar to that obtained by MD simulations of Limbach y Holm \cite{LH1}-\cite{LHK2}

\section{Conclusions}
With the simple technique described here, we were able to reproduce the complex structure of model polyelectrolytes that fare very well with those predicted by the more sophisticated Molecular Dynamic and Monte Carlo simulations{\cite{DRO1,LH1}}.
We even predict situations with single conglomerates and with pearl necklace type conglomerates. We thus show the potentiality of the Celular Automata in the simulation of the trends in the formation of the various types of spatial conformations of polyelectrolytes. We remark on the importance of the charge distribution once the fractional charge is fixed.

\section{Acnowledgement}
This work was supported in part by the grant 04-005-01 from the Decanato de
Investigaci\'on of Universidad Nacional Experimental del T\'achira, in part by
the grant G-9700741 of FONACYT, and in part by the grant C-1279-0402-B from
Consejo de Desarrollo Cient\'{\i}co Humanstico y Tecnol\'ogico of Universidad de
Los Andes.\\
 Numerical calculations were carried out at the computer center 
CeCalcULA.\\

\end{document}